\documentclass[preprint,preprintnumbers,showpacs,aps,prd,amssymb]{revtex4-1}

\usepackage{graphicx}
\usepackage{bm}
\usepackage{amsmath}
\def\calB{{\cal B}}
\def\calL{{\cal L}}
\def\calO{{\cal O}}

\def\Ds{{D^{(*)}}}
\def\RDs{{R(D^{(*)})}}
\def\RtDs{{R_\tau (D^{(*)})}}

\def\SM{{\rm SM}}

\def\nn{\nonumber}

\begin{document}
\title{$B\to D^{(*)}\tau\nu_\tau$ in the 2HDM with an anomalous $\tau$ coupling}
\author{Jong-Phil Lee}
\email{jongphil7@gmail.com}
\affiliation{Sang-Huh College,
Konkuk University, Seoul 05029, Korea}

\begin{abstract}
The puzzle of $\RDs$ associated with $B\to\Ds\tau\nu$ decay is addressed in the two-Higgs-doublet model.
An anomalous coupling of $\tau$ to the charged Higgs is introduced to fit the data from BaBar, Belle, and
LHCb. 
It is shown that all four types of the model yield similar values for the minimum $\chi^2$.
We also show that the newly normalized $\RDs$ with the branching ratio of $B\to\tau\nu$ decay exhibits
a much smaller minimum $\chi^2$.
\end{abstract}

\maketitle
%%%%%%%%%%%%%%%%%%%%%%%%%%%%%%%%%%%%%%%%%%%%%%%%%%%%%%%%%%%%%%%%%%%%%%%%%%%%%%%%
\section{Introduction}
%%%%%%%%%%%%%%%%%%%%%%%%%%%%%%%%%%%%%%%%%%%%%%%%%%%%%%%%%%%%%%%%%%%%%%%%%%%%%%%%
One of the most interesting puzzles in flavor physics in recent years has been the excess of the semitaunic $B$ decays,
$B\to\Ds\tau\nu_\tau$.
The excess is well expressed in terms of the ratio
\begin{equation}
\RDs\equiv\frac{\calB(B\to\Ds\tau\nu)}{\calB(B\to\Ds\ell\nu)}~,
\end{equation}
where $\calB$ is the branching ratio.
The standard model (SM) prediction is \cite{Na,Fajfer}
\begin{eqnarray}
\label{RSM}
R(D)_\SM&=&0.300\pm0.008~,\nn\\
R(D^*)_\SM&=&0.252\pm0.003~.
\end{eqnarray}
The BaBar Collaboration has reported that the measured $R(D)$ exceeds the SM prediction by $2.0\sigma$,
while $R(D^*)$ exceeds the SM prediction by $2.7\sigma$, and 
the combined significance of the disagreement is $3.4\sigma$ \cite{BaBar1,BaBar_PRL}.
BaBar analyzed the possible effect of a charged Higgs boson in the Type-II two-Higgs-doublet model (2HDM),
and excluded the model at the 99.8\% confidence level.
\par
The Belle measurements of $\RDs$ are slightly smaller than those of BaBar, 
but still larger than the SM expectations \cite{Belle1,Belle1607}.
Interestingly, Belle's results are compatible with the Type-II 2HDM in the $\tan\beta/m_{H^\pm}$ region 
around $0.45c^2/\text{GeV}$ (where $\beta$ is the ratio of the two vacuum expectation values of the 2HDM) and zero \cite{Belle1}, and
recent measurements of $R(D^*)$ are consistent with the SM predictions \cite{Belle1612}
On the other hand, LHCb reported that $R(D^*)$ is larger than the SM predictions by $2.1\sigma$ \cite{LHCb1}.
\par
In this paper we try to fit the global data on $\RDs$ with the 2HDM of all types.
The 2HDM is a natural extension of the SM Higgs sector, so it has been tested to fit the $\RDs$ puzzle
\cite{Andreas,Fazio,Cline,Koerner}.
Out of all the types of 2HDM, the Type-II model is the most promising because the new physics (NP) effects
are involved with the coupling of $\tan^2\beta$ while for other types the couplings are 1 or $1/\tan^2\beta$.
As mentioned before, there is tension between BaBar and Belle regarding the compatibility 
of the Type-II 2HDM to the data.
In this analysis we introduce an anomalous $\tau$ coupling to the charged Higgs \cite{Dhargyal}.
Since the NP effects are enhanced by new couplings and suppressed by the charged Higgs mass,
the new couplings should be large enough to allow a heavy charged Higgs to fit the data.
We also investigate possible roles of leptonic decay $B\to\tau\nu$ to solve the $\RDs$ puzzle.
It was suggested that the normalized $\RDs$ with $\calB(B\to\tau\nu)$, $\RtDs$ are consistent with the SM \cite{Nandi}.
We implement the global $\chi^2$ fitting to $\RDs$ as well as $\RtDs$ with the anomalous $\tau$ coupling.
\par
The paper is organized as follows.
Section II introduces the 2HDM with the anomalous $\tau$ coupling to describe
$B\to\Ds\tau\nu$ and $B\to\tau\nu$ transitions.
In Sec.\ III $\RDs$ and $\RtDs$ are expressed in the 2HDM with the new coupling.
Our results and discussions are given in Sec.\ IV, and conclusions follow in Sec.\ V.

%%%%%%%%%%%%%%%%%%%%%%%%%%%%%%%%%%%%%%%%%%%%%%%%%%%%%%%%%%%%%%%%%%%%%%%%%%%%%%%%
\section{2HDM with anomalous $\tau$ couplings}
%%%%%%%%%%%%%%%%%%%%%%%%%%%%%%%%%%%%%%%%%%%%%%%%%%%%%%%%%%%%%%%%%%%%%%%%%%%%%%%%
The Yukawa interaction in the 2HDM is given by \cite{Aoki}
\begin{eqnarray}
\label{LY}
\calL_{\rm Yukawa}&=&
-\sum_{f=u,d,\ell}\frac{m_f}{v}\left(
\xi_h^f{\bar f}fh + \xi_H^f{\bar f}fH -i\xi_A^f{\bar f}\gamma_5 fA\right)\nn\\
&&\left[
 \frac{\sqrt{2}V_{ud}}{v}{\bar u}\left(m_u\xi_A^uP_L\ + m_d\xi_A^dP_R\right)dH^+
+\frac{\sqrt{2}\xi_A^\ell m_\ell}{v}{\bar\nu}_L\ell_RH^+ +{\rm h.c.}\right]~,
\end{eqnarray}
where $v=\sqrt{v_1^2+v_2^2}=246$ GeV, $v_{1,2}$ are the vacuum expectation values (VEVs) of the scalar fields
$\Phi_{1,2}$ of the 2HDM with $\tan\beta\equiv v_2/v_1$, 
and the $\xi$s are the couplings defined in Table \ref{T1}.
%-------------------- Table 1---------------------------------------------------
\begin{table}
\begin{tabular}{|c||rrr|}\hline
 & $\xi_A^u$ & $\xi_A^d$ & $\xi_A^\ell$ \\\hline\hline
Type-I  & $\cot\beta$ & $-\cot\beta$ & $-\cot\beta$  \\
Type-II & $\cot\beta$ & $ \tan\beta$ & $ \tan\beta$  \\
Type-X & $\cot\beta$ & $-\cot\beta$ & $ \tan\beta$  \\
Type-Y & $\cot\beta$ & $ \tan\beta$ & $-\cot\beta$  \\\hline
\end{tabular}
\caption{$\xi$s for each type of 2HDM.}
\label{T1}
\end{table}
%---------------------------------------------------------------------------------
Here we introduce an anomalous factor $\eta$ to enhance $\xi_A^\tau$ \cite{Dhargyal}.
The motivation is that $\tau$ is screened from the second Higgs VEV $v_2$ and the neutral component of $\Phi_2$
by a factor of $\eta$.
In this case the tau mass is $\sim Y_{\rm Yukawa}v_2/\eta$, effectively enhancing the Yukawa coupling of $\tau$ to $H^\pm$,
while that of $\tau$ to neutral Higgses remains unchanged. 
%%%%%%%%%%%%%%%%%%%%%%%%%%%%%%
This kind of model can be easily constructed within extra dimensions.
For example, as in Refs. \cite{Archer,Agashe}, the overlappings between the wave functions of $\tau$ and the neutral component of $\Phi_2$
over the extra dimension would determine the strength of the $\tau$ coupling to the neutral part of $\Phi_2$.
We could simply assume that the overlapping of $\tau$ and the neutral $\Phi_2$ is rather weak compared to other cases.
%%%%%%%%%%%%%%%%%%%%%%%%%%%%%%
The enhancement occurs for Type-I and Type-Y models because in these models leptons couple only to $\Phi_2$.
The same thing could happen for $\tau$ and the neutral component of $\Phi_1$ to screen $\tau$ from $v_1$,
resulting in $\eta$ enhancement for $\tau$-$H^\pm$ couplings in Type-II and X models.
In this work we assume that phenomenologically $\tau$ couplings to $H^\pm$ are enhanced
by a factor of $\eta$ for all types of the model,
\begin{equation}
\xi_A^\tau \to \eta\xi_A^\tau~.
\end{equation}
Now the effective Lagrangian for the $b\to c\ell\nu$ transition is
\begin{eqnarray}
\label{Lb2c}
\calL(b\to c\ell\nu)&=&\calL(b\to c\ell\nu)_{\rm SM}+\calL(b\to c\ell\nu)_{\rm 2HDM}\\\nn
&=&
\frac{G_FV_{cb}}{\sqrt{2}}\left[{\bar c}\gamma^\mu\left(1-\gamma_5\right)b\
{\bar\ell}\gamma_\mu\left(1-\gamma_5\right)\nu_\ell\right]\\\nn
&&
+\frac{V_{cb}}{m_{H^\pm}^2}\left[{\bar c}\left(g_s^c+g_p^c\gamma_5\right)b\
{\bar\ell}\left(f_s^\ell - f_p^\ell\gamma_5\right)\nu_\ell\right]~,
\end{eqnarray}
where 
\begin{eqnarray}
\label{gcfl}
g_s^c &=& \frac{m_c\xi_A^u +m_b\xi_A^d}{\sqrt{2}v}~,~~~
g_p^c =      \frac{-m_c\xi_A^u +m_b\xi_A^d}{\sqrt{2}v}~,\\
f_s^\ell &=& f_p^\ell = -\frac{m_\ell\xi_A^\ell}{\sqrt{2}v}~.
\end{eqnarray}
For $B\to\tau\nu$ decay, 
\begin{eqnarray}
\label{Lb2u}
\calL(B\to\tau\nu)&=&\calL(B\to\tau\nu)_{\rm SM}+\calL(B\to\tau\nu)_{\rm 2HDM}\\\nn
&=&
\frac{G_FV_{ub}}{\sqrt{2}}\left[{\bar u}\gamma^\mu\left(1-\gamma_5\right)b\
{\bar\tau}\gamma_\mu\left(1-\gamma_5\right)\nu_\tau\right]\\\nn
&&
+\frac{V_{ub}}{m_{H^\pm}^2}\left[{\bar u}\left(g_s^u+g_p^u\gamma_5\right)b\
{\bar\tau}\left(f_s^\tau - f_p^\tau\gamma_5\right)\nu_\tau\right]~,
\end{eqnarray}
where 
\begin{eqnarray}
\label{gcfl}
g_s^u &=& \frac{m_u\xi_A^u +m_b\xi_A^d}{\sqrt{2}v}~,~~~
g_p^u =      \frac{-m_u\xi_A^u +m_b\xi_A^d}{\sqrt{2}v}~,\\
f_s^\tau &=& f_p^\tau = -\frac{m_\tau\xi_A^\tau}{\sqrt{2}v}~.
\end{eqnarray}
Note that $\xi_A^\tau$ contains the enhancement factor $\eta$,
$\xi_A^\tau=\eta\xi_A^{\ell=e,\mu}$.

%%%%%%%%%%%%%%%%%%%%%%%%%%%%%%%%%%%%%%%%%%%%%%%%%%%%%%%%%%%%%%%%%%%%%%%%%%%%%%%%
\section{$B\to\Ds\tau\nu$ and $B\to\tau\nu$ decays}
%%%%%%%%%%%%%%%%%%%%%%%%%%%%%%%%%%%%%%%%%%%%%%%%%%%%%%%%%%%%%%%%%%%%%%%%%%%%%%%%

The decay rates of $B\to\Ds\ell\nu$ in the 2HDM can be expressed as
\begin{equation}
\Gamma^\Ds = \Gamma^\Ds_\SM +\Gamma^\Ds_{\rm mix} +\Gamma^\Ds_{H^\pm}~.
\end{equation}
The differential decay rates for $B\to D\ell\nu$ are given by
\begin{eqnarray}
\label{dGDdsSM}
\frac{d\Gamma^D_\SM}{ds}&=&
\frac{G_F^2 |V_{cb}|^2}{96\pi^3m_B^2}\left\{
 4m_B^2P_D^2\left(1+\frac{m_\ell^2}{2s}\right)|F_1|^2 \right. \\\nn
&&
\left. +m_B^4\left(1-\frac{m_D^2}{m_B^2}\right)^2\frac{3m_\ell^2}{2s}|F_0|^2\right\}
 \left(1-\frac{m_\ell^2}{s}\right)^2P_D~,
\\
\label{dGDdsmix}
\frac{d\Gamma^D_{\rm mix}}{ds}&=&
\frac{G_F}{\sqrt{2}m_{H^\pm}^2}\frac{g_s^c|V_{cb}|^2}{32\pi^3}\left(f_s^\ell+f_p^\ell\right)m_\ell\\\nn
&&\times
\left(1-\frac{m_D^2}{m_B^2}\right)\left(\frac{m_B^2-m_D^2}{m_b-m_c}\right)|F_0|^2
\left(1-\frac{m_\ell^2}{s}\right)^2 P_D~,
\\
\label{dGDdsHpm}
\frac{d\Gamma^D_{H^\pm}}{ds}&=&
\frac{(g_s^c)^2|V_{cb}|^2}{64\pi^3 m_{H^\pm}^4 m_B^2}\left[(f_s^\ell)^2+(f_p^\ell)^2\right]
|F_0|^2 s\left(1-\frac{m_\ell^2}{s}\right)^2 P_D~,
\end{eqnarray}
where $s=(p_B-p_D)^2$ is the momentum-transfer squared, and
\begin{equation}
\label{PD}
P_D\equiv\frac{\sqrt{s^2+m_B^4+m_D^4-2(sm_B^2+sm_D^2+m_B^2m_D^2)}}{2m_B}~,
\end{equation}
is the momentum of $D$ in the $B$ rest frame.
The form factors $F_0$ and $F_1$ are given by
\begin{eqnarray}
\label{F01}
F_0 &=& \frac{\sqrt{m_Bm_D}}{m_B+m_D} (w+1)S_1~,\\
F_1 &=& \frac{\sqrt{m_Bm_D}(m_B+m_D)}{2m_BP_D}\sqrt{w^2-1}V_1~,
\end{eqnarray}
where
\begin{eqnarray}
V_1(w)&=&
V(1)\left[1-8\rho_D^2z(w)+(51\rho_D^2-10)z(w)^2-(252\rho_D^2-84)z(w)^3\right]~,\\
S_1(w)&=&
V_1(w)\left\{
1+\Delta\left[-0.019+0.041(w-1)-0.015(w-1)^2\right]\right\}~,
\end{eqnarray}
with
\begin{eqnarray}
w&=&\frac{m_B^2+m_D^2-s}{2m_Bm_D}~,~~~
z(w)=\frac{\sqrt{w+1}-\sqrt{2}}{\sqrt{w+1}+\sqrt{2}}~,\\
\rho_D^2&=&1.186\pm0.055~,~~~
\Delta=1\pm1~.
\end{eqnarray}
%%%%%%%%%%%%
\par
For $B\to D^*\ell\nu$ decay,
\begin{eqnarray}
\label{dGDsds}
\frac{d\Gamma^{D^*}_\SM}{ds} 
&=& 
\frac{G_F^2|V_{cb}|^2}{96\pi^3m_B^2}\left[
  \left( |H_+|^2+|H_-|^2+|H_0|^2\right)\left(1+\frac{m_\ell^2}{2s}\right)
  +\frac{3m_\ell^2}{2s}|H_s|^2\right] \\\nn
&&\times
  s\left(1-\frac{m_\ell^2}{s}\right)^2P_{D^*}~,\\
\frac{d\Gamma^{D^*}_{\rm mix}}{ds} 
&=& 
\frac{G_F}{\sqrt{2}}\frac{m_\ell g_p^c|V_{cb}|^2}{8\pi^3 m_{H^\pm}^2}
\frac{f_s^\ell+f_p^\ell}{m_b+m_c}A_0^2
\left(1-\frac{m_\ell^2}{s}\right)^2P_{D^*}^3~,\\
\frac{d\Gamma^{D^*}_{H^\pm}}{ds} 
&=&
\frac{(g_p^c)^2|V_{cb}|^2}{16\pi^3 m_{H^\pm}^4}
\frac{(f_s^\ell)^2+(f_p^\ell)^2}{(m_b+m_c)^2} A_0^2
s\left(1-\frac{m_\ell^2}{s}\right)^2P_{D^*}^3~,
\end{eqnarray}
where $P_{D^*}=P_D(m_D\to m_{D^*})$.
The form factors are given by
\begin{eqnarray}
\label{Hpm0s}
H_\pm(s) &=&
(m_B+m_{D^*})A_1(s)\mp\frac{2m_B}{m_B+m_{D^*}}P_{D^*} V(s)~,\\
H_0(s) &=&
\frac{-1}{2m_{D^*}\sqrt{s}}\left[
\frac{4m_B^2P_{D^*}^2}{m_B+m_{D^*}}A_2(s)
-(m_B^2-m_{D^*}^2-s)(m_B+m_{D^*})A_1(s)\right]~,\\
H_s(s) &=&
\frac{2m_B P_{D^*}}{\sqrt{s}} A_0(s)~,
\end{eqnarray}
where
\begin{eqnarray}
\label{A012V}
A_1(w^*) &=& \frac{w^*+1}{2}r_{D^*}h_{A_1}(w^*)~,\\
A_0(w^*) &=& \frac{R_0(w^*)}{r_{D^*}}h_{A_1}(w^*)~,\\
A_2(w^*) &=& \frac{R_2(w^*)}{r_{D^*}}h_{A_1}(w^*)~,\\
V(w^*)    &=& \frac{R_1(w^*)}{r_{D^*}}h_{A_1}(w^*)~,\\
\end{eqnarray}
with
\begin{equation}
\label{wsrDs}
w^* = \frac{m_B^2+m_{D^*}^2-s}{2m_Bm_{D^*}}~,~~~
r_{D^*} = \frac{2\sqrt{m_Bm_{D^*}}}{m_B+m_{D^*}}~,
\end{equation}
and
\begin{eqnarray}
\label{hR}
h_{A_1}(w^*)&=&
h_{A_1}(1)\left[1-8\rho_{D^*}^2z(w^*)+(53\rho_{D^*}^2-15)z(w^*)^2
-(231\rho_D{^*}^2-91)z(w^*)^3\right]    ~,\\
R_0(w^*) &=&
R_0(1)-0.11(w^*-1)+0.01(w^*-1)^2    ~,\\
R_1(w^*) &=&    
R_1(1)-0.12(w^*-1)+0.05(w^*-1)^2    ~,\\
R_2(w^*) &=&    
R_2(1)+0.11(w^*-1)-0.01(w^*-1)^2     ~.
\end{eqnarray}
Here\cite{Dhargyal}
\begin{eqnarray}
\rho_{D^*}^2 &=& 1.207\pm0.028~,~~~ R_0(1)=1.14\pm 0.07~,\\
R_1(1)&=&1.401\pm0.033~,~~~ R_2(1)=0.854\pm 0.020~.
\end{eqnarray}
%
%%%%%%%%%%%%%%%%%%%%%%%%%%%%%
%
For the leptonic two-body decay $B\to\tau\nu$, the branching ratio is 
\begin{equation}
\calB(B\to\tau\nu)=
\calB(B\to\tau\nu)_\SM\left(1+ r_{H^\pm}\right)^2~,
\end{equation}
where
\begin{eqnarray}
\calB(B\to\tau\nu)_\SM &=&
\frac{G_F^2 |V_{ub}|^2m_\tau^2 m_B}{8\pi}f_B^2
\left(1-\frac{m\tau^2}{m_B^2}\right)^2\tau_B~,\\
r_{H^\pm}&=&
\frac{(m_u/m_b)\xi_A^u-\xi_A^d}{1+m_u/m_b}\xi_A^\tau\left(\frac{m_B}{m_{H^\pm}}\right)^2
~.
\end{eqnarray}
Here $f_B$ and $\tau_B$ are the decay constant and lifetime of $B$, respectively. 
\par
The experimental data are summarized in Table \ref{T2} \cite{Nandi}.
%-------------------- Table 2---------------------------------------------------
\begin{table}
\begin{tabular}{|c|| lll |}\hline
              & ~$R(D)$ & ~$R(D^*)$ & ~$\calB(B\to\tau\nu)$ \\\hline\hline
 BABAR & ~$0.440\pm0.058\pm0.042$ & ~$0.332\pm0.024\pm0.018$ \cite{BaBar1} & ~$1.83^{+0.53}_{-0.49}\times 10^{-4}$ \cite{BaBar2}\\
 Belle(2015) & ~$0.375\pm0.064\pm0.026$ & ~$0.293\pm0.038\pm0.015$ \cite{Belle1}  & ~$(1.25\pm0.28)\times 10^{-4}$ \cite{Belle3}\\
 Belle(1607) & ~$-$ & ~$0.302\pm0.030\pm0.011$ \cite{Belle1607}& ~$-$ \\
 Belle(1612) & ~$-$ & ~$0.270\pm0.035^{+0.028}_{-0.025}$ \cite{Belle1612}& ~$-$ \\
 LHCb          & ~$-$ & ~$0.336\pm0.027\pm0.030$ \cite{LHCb1} & ~$-$ \\\hline
 \end{tabular}
\caption{Experimental data for $R(\Ds)$ and $\calB(B\to\tau\nu)$.
For $\RDs$ measurements the uncertainties are $\pm$(statistical)$\pm$(systematic).}
\label{T2}
\end{table}
%---------------------------------------------------------------------------------
At first we try to fit the data of Table \ref{T2} by minimizing $\chi^2$.
BABAR results \cite{BaBar1} already ruled out the Type-II 2HDM.
We introduce an anomalous $\tau$ coupling for all types of 2HDM,
which will be shown to significantly reduce the $\chi^2$ minimum.
\par
In addition, it was suggested that the ratio
\begin{equation}
R_\tau(\Ds)\equiv\frac{R(\Ds)}{\calB(B\to\tau\nu)}~,
\label{RtDs}
\end{equation}
has some advantages in this analysis \cite{Nandi}.
First of all the $\tau$ detection systematics is canceled in the ratio.
But it should be noted that the ratio of Eq.\ (\ref{RtDs}) introduces the theoretical error on $V_{ub}$.
%-------------------- Table 3---------------------------------------------------
\begin{table}
\begin{tabular}{|c|| ll |}\hline
              		        & ~$R_\tau(D)\times 10^3$ & ~$R_\tau(D^*)\times 10^3$  \\\hline\hline
 BABAR 		        & ~$2.404 \pm 0.838$ & ~$1.814 \pm 0.582$  \\
 BABAR($\tau$ tag)  & ~$5.96 \pm 2.26$ & ~$4.49 \pm 3.54$ \\
 Belle(2015) 		& ~$3.0 \pm 1.1$ & ~$2.344 \pm 0.799$  \\
 Belle($\tau$ tag)	& ~$5.7 \pm 3.3$ & ~$4.49 \pm 3.54$ \\
 Belle(1607) 		& ~$-$ & ~$2.416\pm 0.794$  \\
 Belle(1612) 		& ~$-$ & ~$2.160\pm 0.835$  \\\hline
 \end{tabular}
\caption{$R_\tau(\Ds)$ values for different experiments \cite{Nandi}.
The value of "Belle(1612)" is a new one.}
\label{T3}
\end{table}
%---------------------------------------------------------------------------------
We use the values of $R_\tau(\Ds)$ in Table\ \ref{T3} for the fit.
%
%%%%%%%%%%%%%%%%%%%%%%%%%%%%%%%%%%%%%%%%%%%%%%%%%%%%%%%%%%%%%%%%%%%%%%%%%%%%%%%%
\section{Results and Discussions}
\begin{figure}
\begin{tabular}{cc}
\includegraphics[scale=0.12]{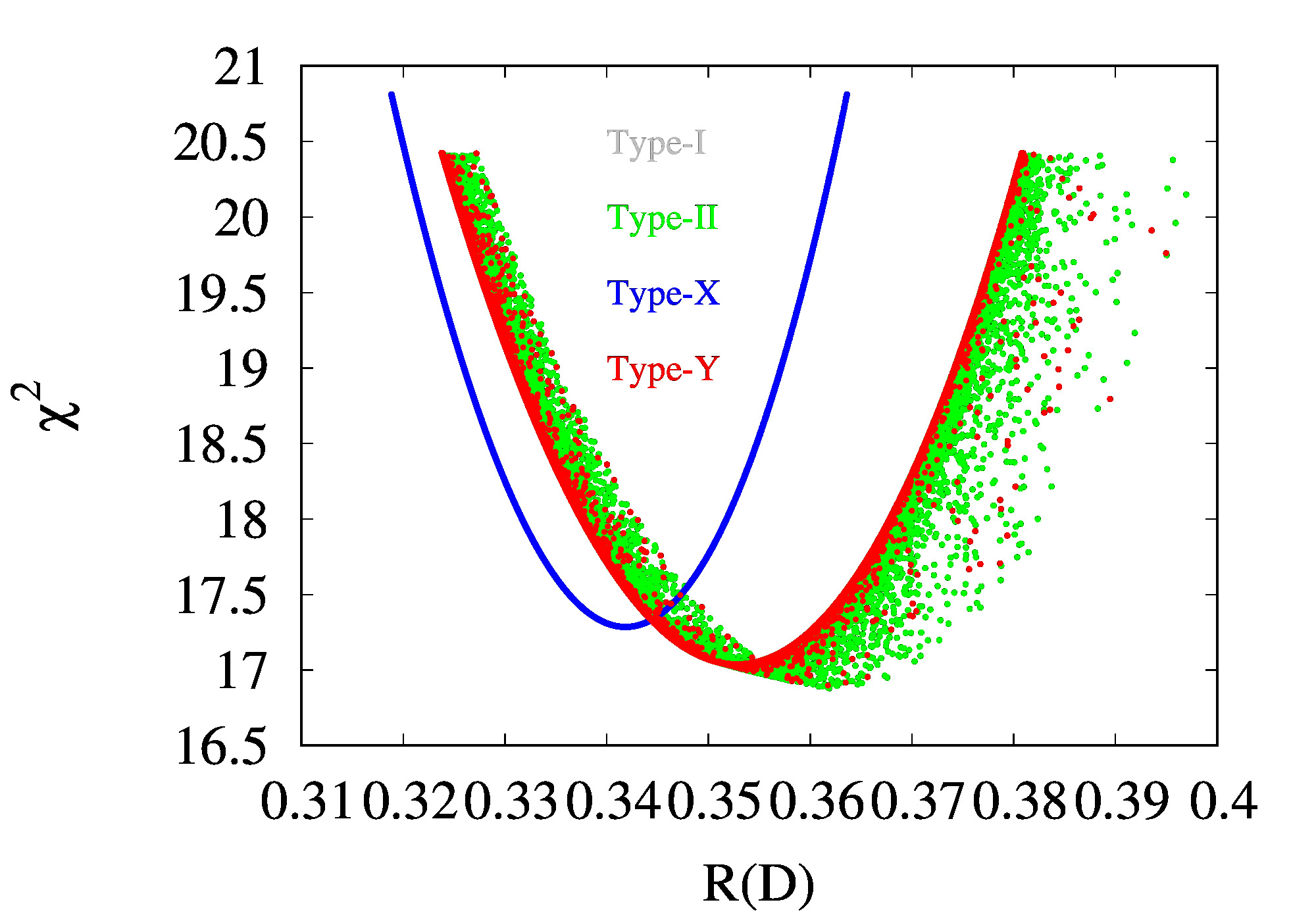} &\includegraphics[scale=0.12]{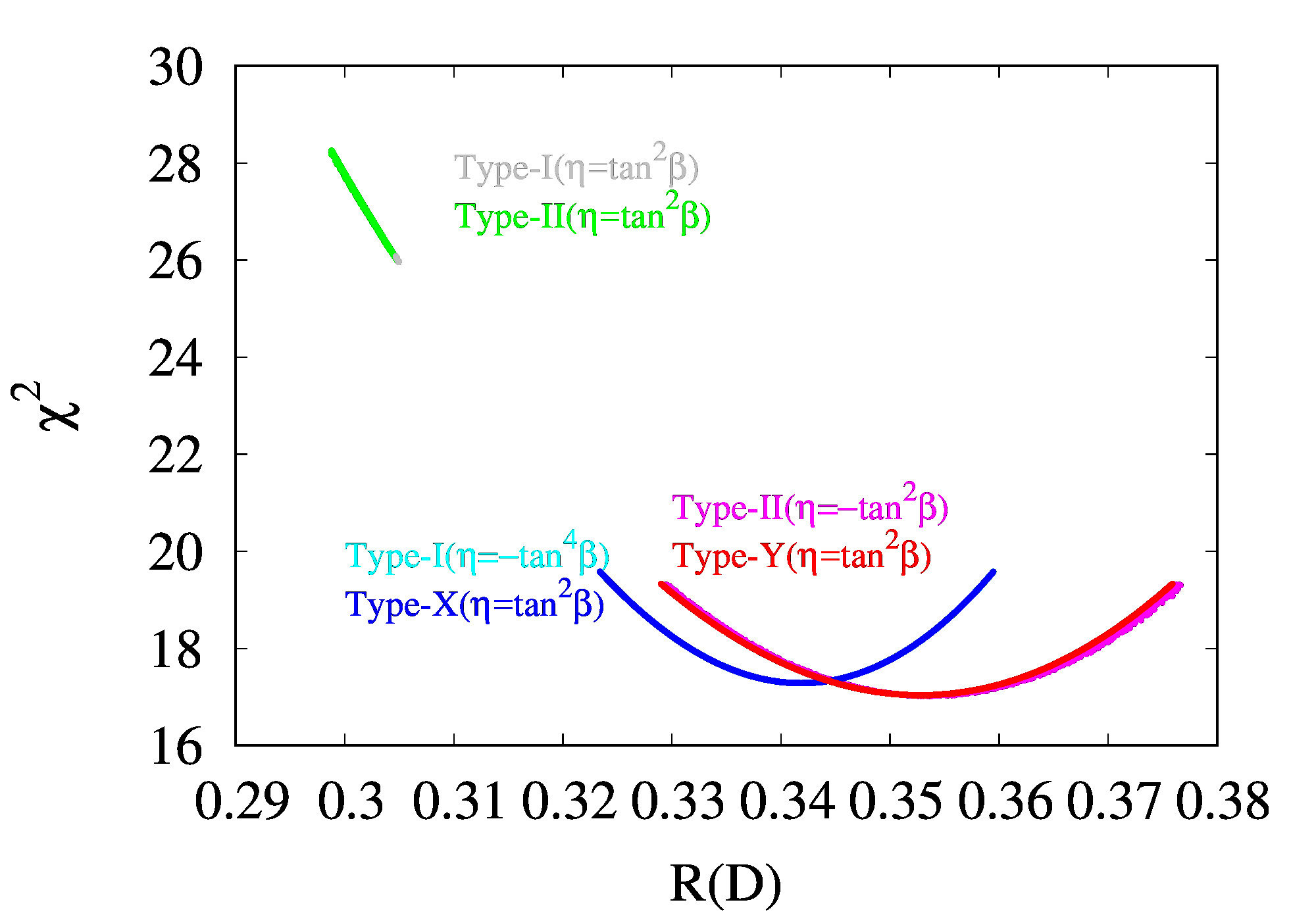}\\
(a) & (b)\\
\end{tabular}
\caption{\label{RD1}
$\chi^2$ vs $R(D)$ for (a) free anomalous couplings and (b) $\eta=\tan^2\beta$ at the $1\sigma$ level.
In panel (b) Type-I with $\eta=-\tan^4\beta$ and Type-II with $\eta=-\tan^2\beta$ are also shown.
In panel (a), Type-I and Type-X are overlapped;
in panel (b), Type-I($\eta=-\tan^4\beta$) and Type-X($\eta=\tan^2\beta$) are overlapped.}
\end{figure}

\begin{figure}
\includegraphics[scale=0.2]{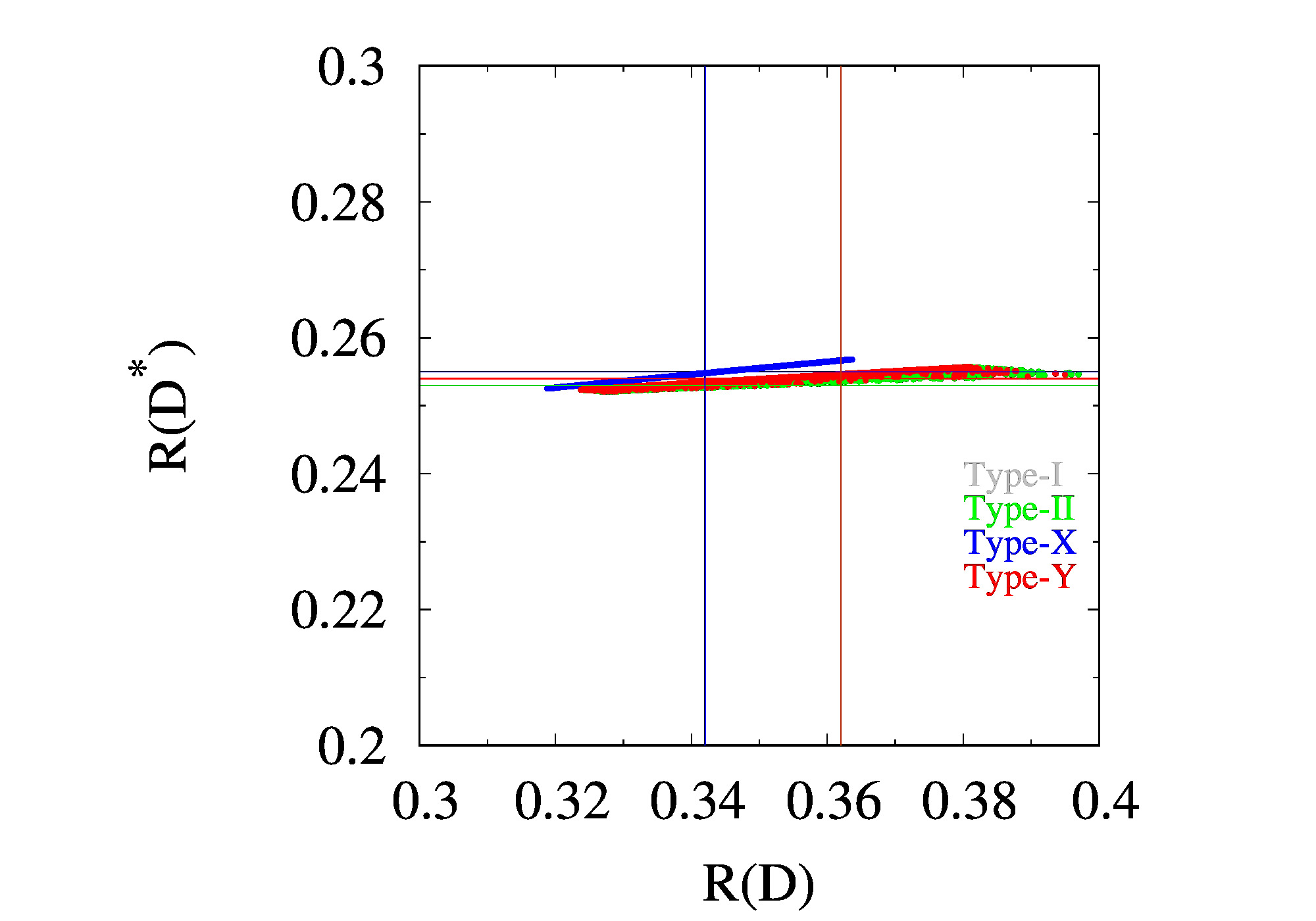}
\caption{\label{RDRDs}
Allowed region in the $R(D)$-$R(D^*)$ plane at the $1\sigma$ level with free $\eta$.
Vertical and horizontal lines are the best-fit points.
}
\end{figure}

%%%%%%%%%%%%%%%%%%%%%%%%%%%%%%%%%%%%%%%%%%%%%%%%%%%%%%%%%%%%%%%%%%%%%%%%%%%%%%%%
In our analysis $\tan\beta$ and $M_{H^\pm}$ are by default the fitting parameters to minimize $\chi^2$, defined by
\begin{equation}
\chi^2=\sum_i\frac{(x_i-\mu_i)^2}{(\delta\mu_i)^2}~,
\end{equation}
where the $x_i$s are model predictions and the $(\mu_i\pm\delta\mu_i)$s are experimental data.
Figure\ \ref{RD1} shows the $R(D)$ values vs $\chi^2$ with the anomalous $\tau$ coupling $\eta$.
In Fig.\ \ref{RD1}(a), $\eta$ is set to be an additional fitting parameter, $-1000\le\eta\le 1000$.
Plots for the Type-I and Type-X models are overlapped.
%----------------- Figure 1 ------------------------------------------------
%------------------------------------------------------------------------------
As can be seen from Eqs.\ (\ref{dGDdsmix}) and (\ref{dGDdsHpm}), the 2HDM contributes to $R(D)$ as
\begin{equation}
\RDs_{H^\pm}\sim
\frac{g_{s,p}^c f_s^\tau}{m_{H^\pm}^2}
+\left(\frac{g_{s,p}^c f_s^\tau}{m_{H^\pm}^2}\right)^2 ~,
\label{RDHpm}
\end{equation}
where the coefficients are omitted for simplicity.
For free $\eta$, Types I and X behave similarly because $\xi_A^u$ and $\xi_A^d$ are the same (see Table\ \ref{T1}).
This is also true for Types II and Y. 
We also consider the case of fixed $\eta\equiv\tan^2\beta$ as in Ref. \cite{Dhargyal} in Fig.\ \ref{RD1}(b).
The dominant contribution to Eq.\ (\ref{RDHpm}) comes from
\begin{eqnarray}
\RDs_{H^\pm}&\sim& 
-\frac{(m_b\xi_A^d)(m_\tau\eta\xi_A^\ell)}{2v^2m_{H^\pm}^2}
+\left[\frac{(m_b\xi_A^d)(m_\tau\eta\xi_A^\ell)}{2v^2m_{H^\pm}^2}\right]^2 \nn\\
&=&
\label{RDHpm2}
\begin{cases}
-\left(\frac{m_b m_\tau}{2v^2m_{H^\pm}^2}\right)(\eta\cot^2\beta) 
   + \left[\frac{m_b m_\tau}{2v^2m_{H^\pm}^2}\right]^2(\eta\cot^2\beta)^2 & \text{for Type-I} \\
-\left(\frac{m_b m_\tau}{2v^2m_{H^\pm}^2}\right)(\eta\tan^2\beta)
   + \left[\frac{m_b m_\tau}{2v^2m_{H^\pm}^2}\right]^2(\eta\tan^2\beta)^2 & \text{for Type-II} \\  
-\left(\frac{m_b m_\tau}{2v^2m_{H^\pm}^2}\right)(-\eta)
   + \left[\frac{m_b m_\tau}{2v^2m_{H^\pm}^2}\right]^2(-\eta)^2 & \text{for Type-X} \\  
-\left(\frac{m_b m_\tau}{2v^2m_{H^\pm}^2}\right)(-\eta)
   + \left[\frac{m_b m_\tau}{2v^2m_{H^\pm}^2}\right]^2(-\eta)^2 & \text{for Type-Y}~.    \\
\end{cases}
\end{eqnarray}
Since the first term is negative for Types I and II for $\eta\equiv\tan^2\beta>0$, 
the $\chi^2$ values are very poor compared to those for Types X and Y, as shown in Fig.\ \ref{RD1}(b).
If we allow $\eta= -\tan^4\beta$ for the Type-I model, 
the $\chi^2$ distribution over $R(D)$ overlaps with that for the Type-X model.
Similar things happen for the Type-II model with $\eta=-\tan^2\beta$ and the Type-Y model.
In this case, Eq.\ (\ref{RDHpm}) is not the same for Types II and Y; 
the sign of $\eta$ is more relevant to the $\chi^2$ distribution than the power of $\eta$.
We can see that introducing the anomalous $\tau$ coupling improves the $\chi^2$ fitting, 
and any Type of 2HDM model is as good (or bad) as another.
The best-fit values of $\RDs$ and the corresponding minimum 
$\chi^2$ per degree of freedom (d.o.f.) are given in Table \ref{T4},
%-------------------- Table 4---------------------------------------------------
\begin{table}
\begin{tabular}{|c|| cccc |}\hline
Types	& I 		& II	 	 & X 		  & Y \\\hline\hline
$R(D)$	& $0.342$ & $0.362$ & $0.342$ & $0.362$ \\
$R(D^*)$	& $0.255$ & $0.253$ & $0.255$ & $0.254$ \\
$\chi^2_\text{min}/\text{d.o.f.}$ & $2.881$ & $2.813$ & $2.881$ & $2.861$  \\\hline
 \end{tabular}
\caption{The best-fit $\RDs$ values with free $\eta$ for different Types of the model.}
\label{T4}
\end{table}
%---------------------------------------------------------------------------------
and the allowed region of $R(D)$ and $R(D^*)$ at the $1\sigma$ level is given in Fig.\ \ref{RDRDs}
%----------------- Figure 2 ------------------------------------------------
%------------------------------------------------------------------------------
\par
\begin{figure}
\begin{tabular}{cc}
\includegraphics[scale=0.12]{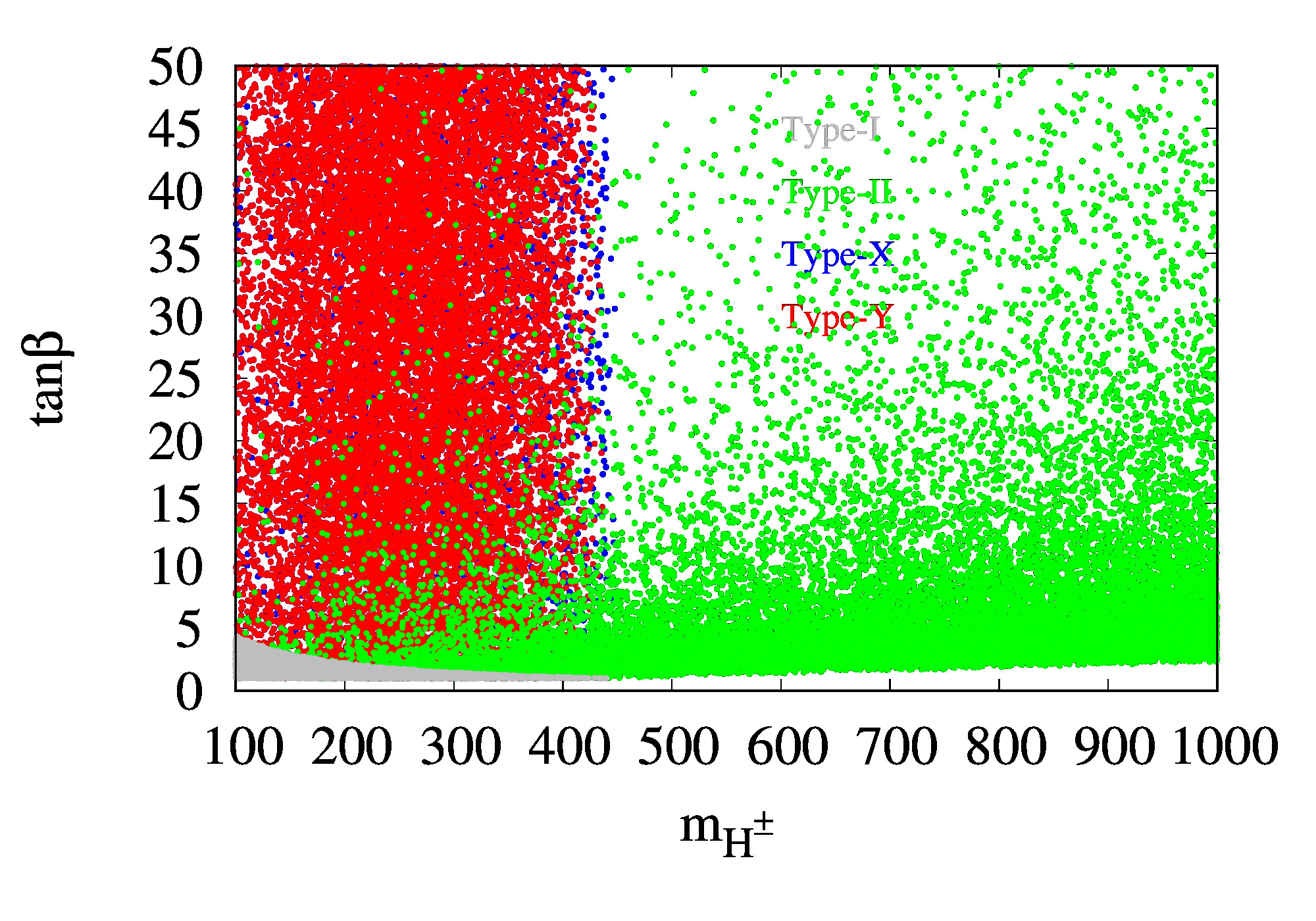}&\includegraphics[scale=0.12]{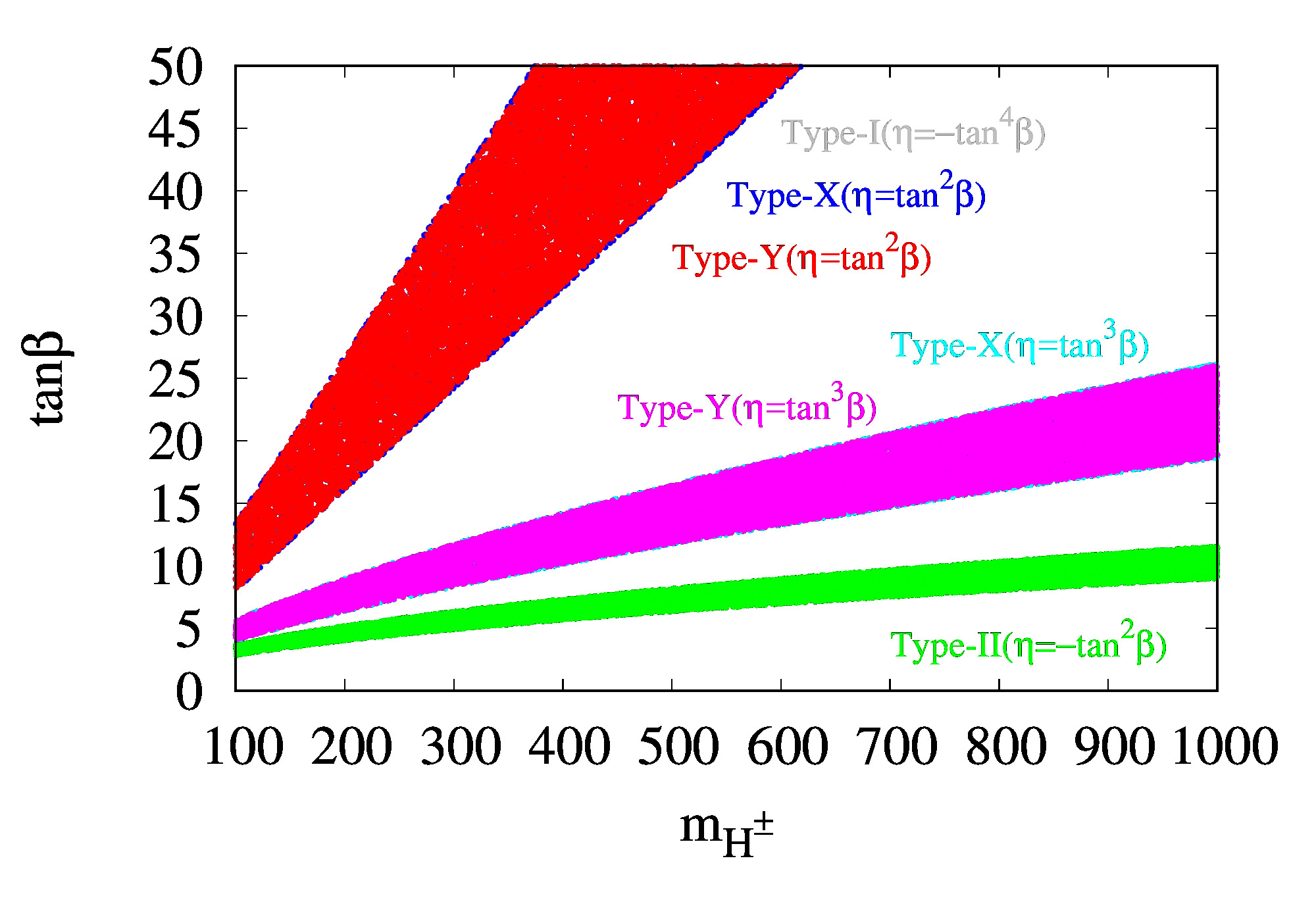}\\
(a) & (b)
\end{tabular}
\caption{\label{RD2}
$\RDs$-fitting results of 
$\tan\beta$ vs $m_{H^\pm}$ for (a) free anomalous couplings within $-1000\le\eta\le 1000$ and 
(b) $\eta=(\pm)\tan^\alpha\beta$ for some $\alpha$ for different Types of the model, at the $1\sigma$ level.
Regions for Type-I with $\eta=-\tan^4\beta$, Type-X with $\eta=\tan^2\beta$,
and Type-Y with $\eta=\tan^2\beta$ are overlapped; 
regions for Type-X with $\eta=\tan^3\beta$ and Type-Y with $\eta=\tan^3\beta$ are also overlapped.
}
\end{figure}

Figure \ref{RD2} shows the allowed region of $m_{H^\pm}$ vs $\tan\beta$.
%----------------- Figure 3 ------------------------------------------------
%------------------------------------------------------------------------------
In Fig.\ \ref{RD2} (a), $\eta$ is a free parameter within $-1000\le\eta\le 1000$.
In this case $m_{H^\pm}$ cannot be large enough because the $R(D)_{H^\pm}$ term of Eq.\ (\ref{RDHpm2}) gets smaller 
and cannot fit the data.
One exception is the Type-II model.
As shown in Eq.\ (\ref{RDHpm2}), there is a $\tan^2\beta$ enhancement for $R(D)_{H^\pm}$, 
which allows $m_{H^\pm}$ to be large.
If we require that the charged Higgs mass is $m_{H^\pm}\gtrsim500$ GeV, 
only the Type-II model survives in Fig.\ \ref{RD2}(a).
In Fig.\ \ref{RD2} (b) we fix $\eta\equiv\pm\tan^\alpha\beta$ for some $\alpha$.
For Types X and Y, the allowed stripe stretches to larger $m_{H^\pm}$ with smaller $\tan\beta$ 
as $\alpha$ goes from 2 to 3.
This is because 
$R(D)_{H^\pm}\sim
\eta m_bm_\tau/m_{H^\pm}^2+\left(\eta m_bm_\tau/m_{H^\pm}^2\right)^2$.
Also shown in Fig.\ \ref{RD2}(b) are the Type-I model with $\eta=-\tan^4\beta$ 
and the Type-II model with $\eta=-\tan^2\beta$ for comparison.
It would be expected from Eq.\ (\ref{RDHpm2}) that stripes for the Type-X and Y models are coincident up to 
$\sim\calO(m_c/m_b)$.
They also overlap with the stripe of the Type-I model with $\eta=-\tan^4\beta$.
The stripe for the Type-II model with $\eta=-\tan^2\beta$ lies in the lowest region of $\tan\beta$ 
since there is already a $\tan^2\beta$ term in $R(D)_{H^\pm}$.
\par
\begin{figure}
\begin{tabular}{cc}
\includegraphics[scale=0.12]{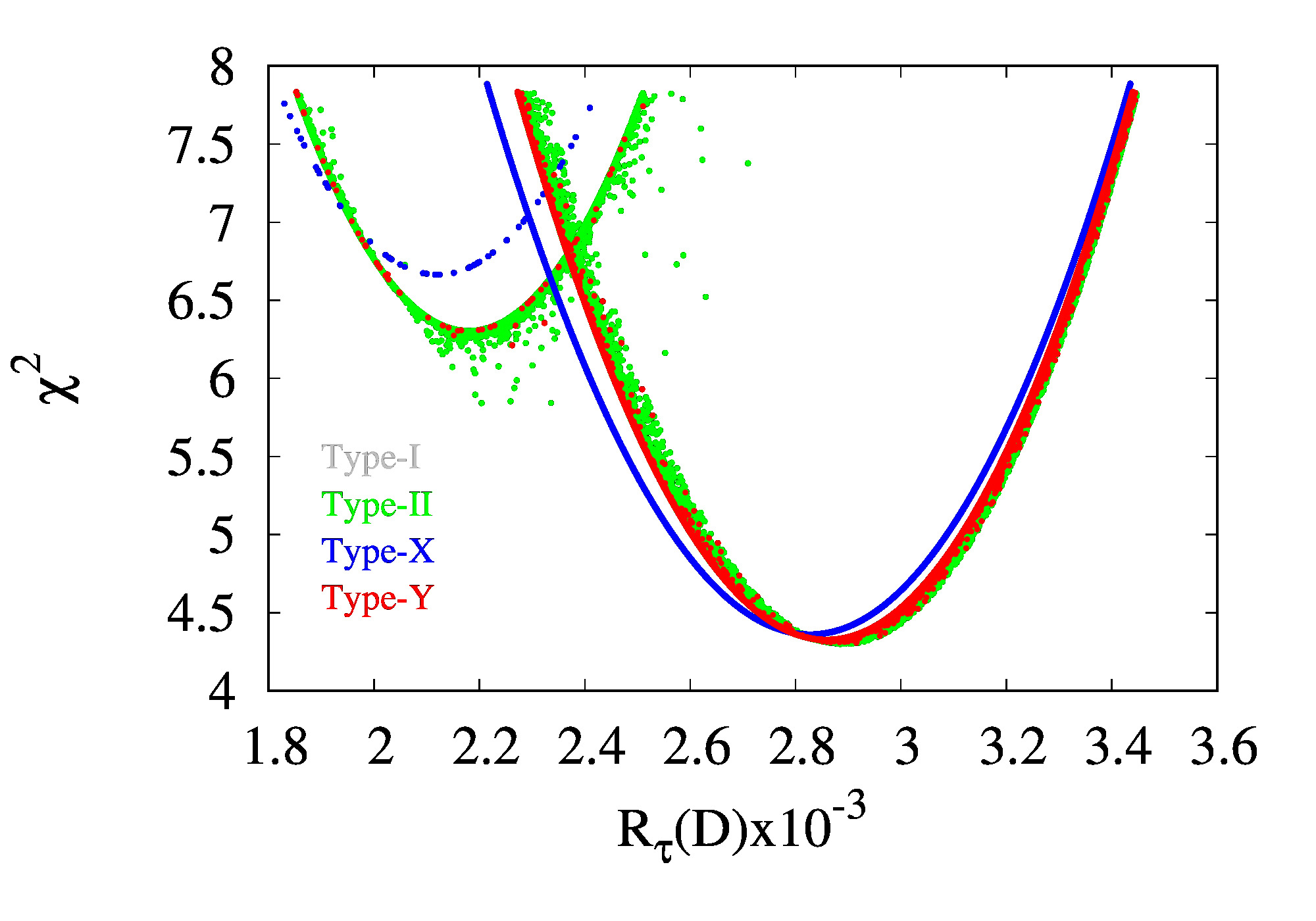}&\includegraphics[scale=0.12]{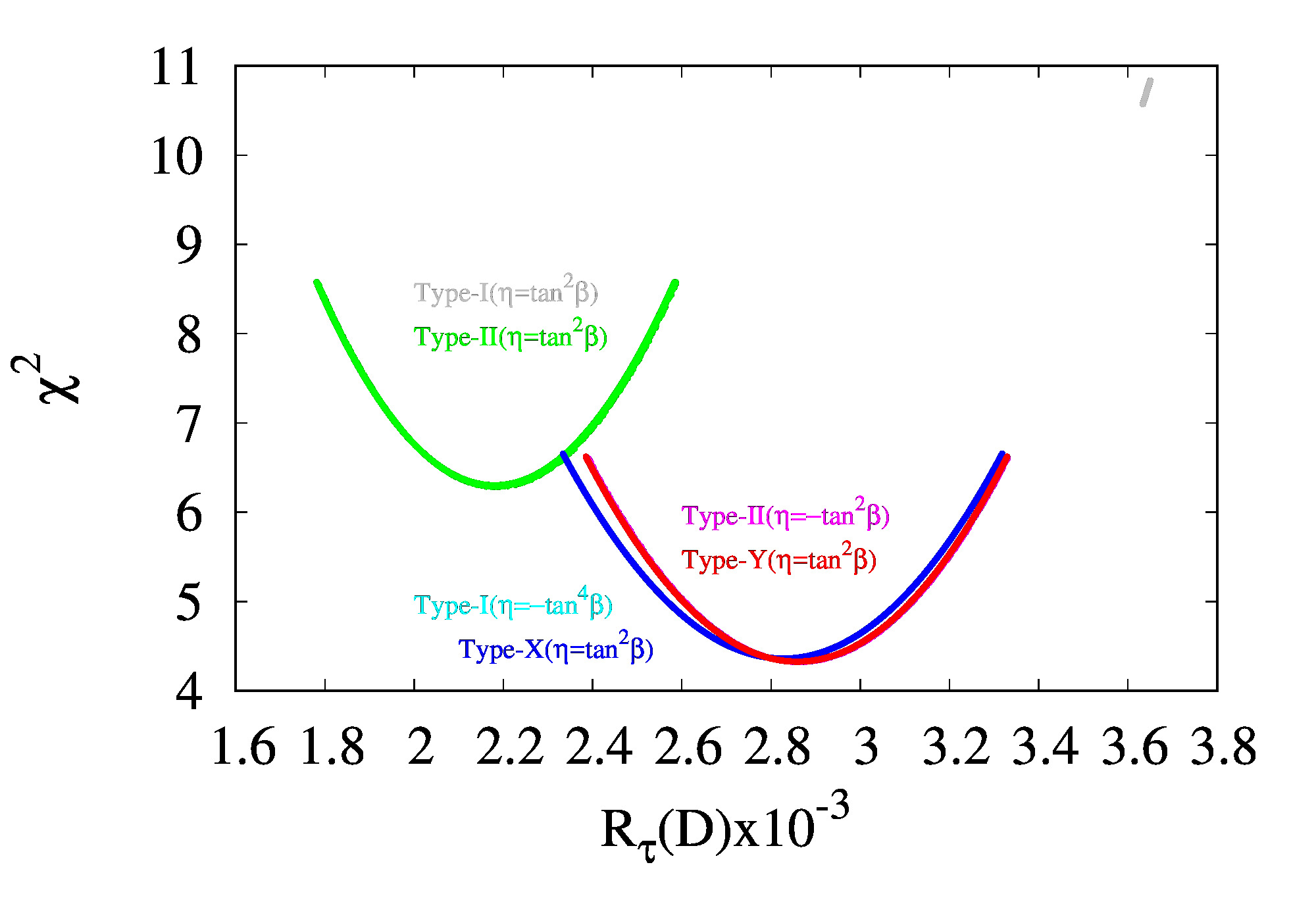}\\
(a) & (b)\\
\end{tabular}
\caption{\label{RtD1}
$\chi^2$ vs $R_\tau(D)$ for (a) free anomalous couplings and (b) $\eta=\tan^2\beta$ at the $1\sigma$ level.
In panel (b) Type-I with $\eta=-\tan^4\beta$ and Type-II with $\eta=-\tan^2\beta$ are also shown.
In panel (a), Type-I is overlapped with Type-X;
in panel (b), grey-green, cyan-blue, and magenta-red curves are overlapped, respectively.}
\end{figure}

%%%%%%%%%%%%%%%%%%%%%%%%%%%%%%%%%%%%%%%%%%%%%
Now we turn to the $\RtDs$.
Figure \ref{RtD1} shows $R_\tau(D)$ vs $\chi^2$.
%----------------- Figure 4 ------------------------------------------------
%------------------------------------------------------------------------------
Note that the minimum $\chi^2$ reduces significantly compared to Fig.\ \ref{RD1};
$\chi^2_\text{min}/\text{d.o.f.}=$ 0.623\ (Type-I, X), 0.614\ (Type-II) 0.615\ (Type-Y) for free $\eta$ 
in Fig.\ \ref{RtD1}(a).  
As discussed in Ref. \cite{Nandi}, $\RtDs$ values from the BABAR and Belle results are consistent with each other
and not so far from the SM predictions \cite{Nandi},
\begin{eqnarray}
R_\tau(D)_\SM&=&(3.136\pm0.628)\times 10^3~,\\
R_\tau(D^*)_\SM&=&(2.661\pm0.512)\times 10^3~.
\end{eqnarray}
\par
In Fig.\ \ref{RtD1}(b) we fix $\eta=\pm\tan^\alpha\beta$ for some $\alpha$.
Any Type of the model is as good as another.
$R_\tau(D^*)$ vs $\chi^2$ shows similar behavior.
The new contribution to $\calB(B\to\tau\nu)$ is 
\begin{equation}
r_{H^\pm} \sim 
-\xi_A^d\xi_A^\tau\left(\frac{m_B}{m_{H^\pm}}\right)^2 =
\label{rHpm}
\begin{cases}
(-\eta\cot^2\beta)\left(\frac{m_B}{m_{H^\pm}}\right)^2 & \text{for Type-I} \\
(-\eta\tan^2\beta)\left(\frac{m_B}{m_{H^\pm}}\right)^2 & \text{for Type-II} \\
\eta\left(\frac{m_B}{m_{H^\pm}}\right)^2 & \text{for Type-X} \\
\eta\left(\frac{m_B}{m_{H^\pm}}\right)^2 & \text{for Type-Y} ~,\\
\end{cases}
\end{equation}
where terms of $\calO(m_u/m_b)$ are neglected.
As in Eq.\ (\ref{RDHpm2}), only the combination of $\xi_A^d\xi_A^\tau$ is relevant, 
and thus the Type-I model with $\eta=-\tan^4\beta$ looks much like the Type-X models 
with $\eta=\tan^2\beta$, and so on.
\par
\begin{figure}
\includegraphics[scale=0.2]{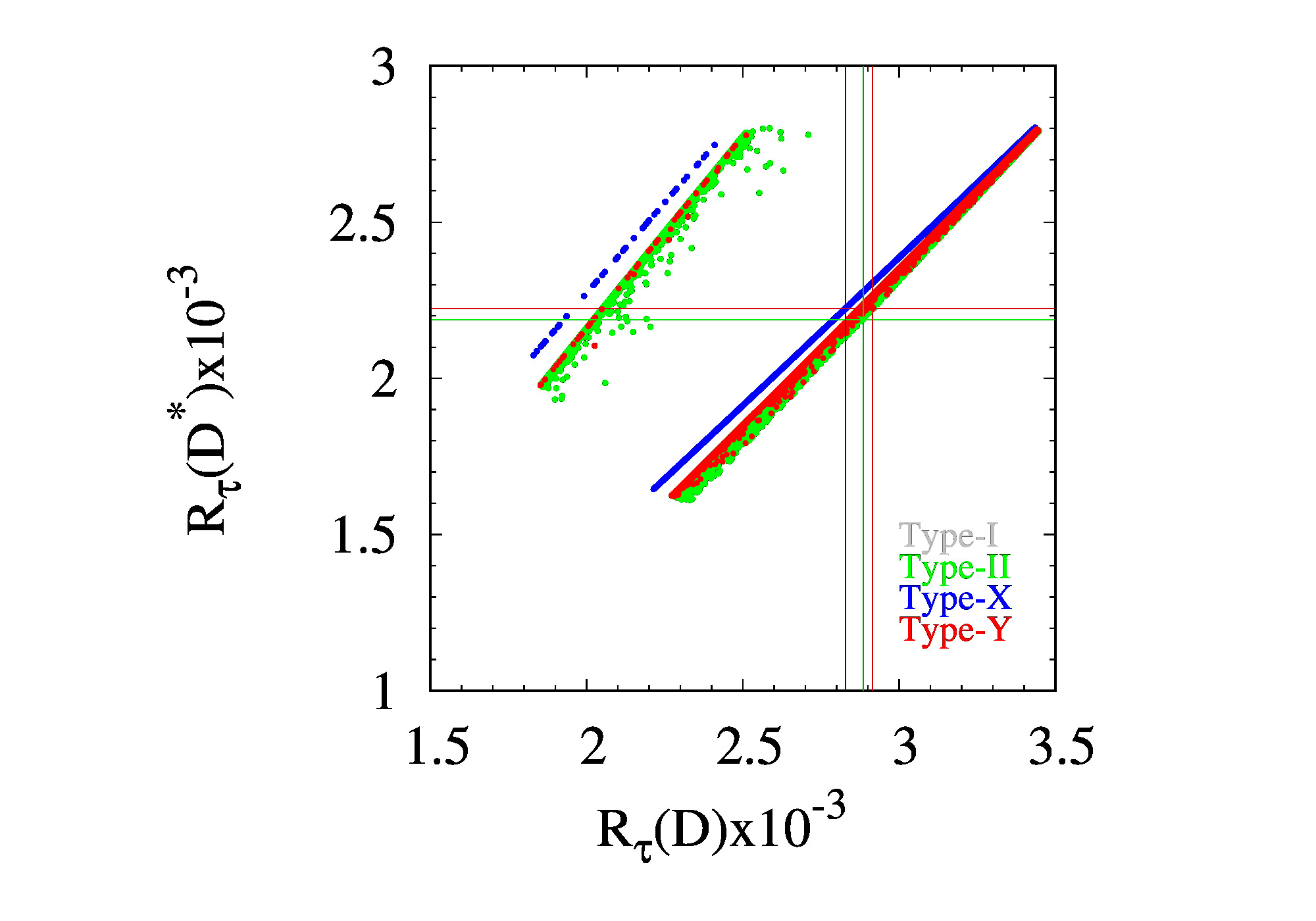}
\caption{\label{RtDRtDs}
Allowed region in the $R_\tau(D)$-$R_\tau(D^*)$ plane at the $1\sigma$ level with free $\eta$.
Vertical and horizontal lines are the best-fit points.
}
\end{figure}

Table \ref{T5} shows the best-fit values of $\RtDs$ and $\chi^2_\text{min}/\text{d.o.f}$,
%-------------------- Table 5---------------------------------------------------
\begin{table}
\begin{tabular}{|c|| cccc |}\hline
Types	& I 		& II	 	 & X 		  & Y \\\hline\hline
$R_\tau(D)\times 10^{-3}$	& $2.828$ & $2.885$ & $2.828$ & $2.915$ \\
$R_\tau(D^*)\times 10^{-3}$	& $2.223$ & $2.188$ & $2.223$ & $2.223$ \\
$\chi^2_\text{min}/\text{d.o.f.}$ & $0.623$ & $0.614$ & $0.623$ & $0.615$  \\\hline
 \end{tabular}
\caption{The best-fit $\RtDs$ values with free $\eta$ for different Types of the model.}
\label{T5}
\end{table}
%---------------------------------------------------------------------------------
and Fig.\ \ref{RtDRtDs} shows the allowed region of $R_\tau(D)$ and $R_\tau(D^*)$ at the $1\sigma$ level.
%----------------- Figure 5 ------------------------------------------------
%------------------------------------------------------------------------------
\par
\begin{figure}
\begin{tabular}{cc}
\includegraphics[scale=0.12]{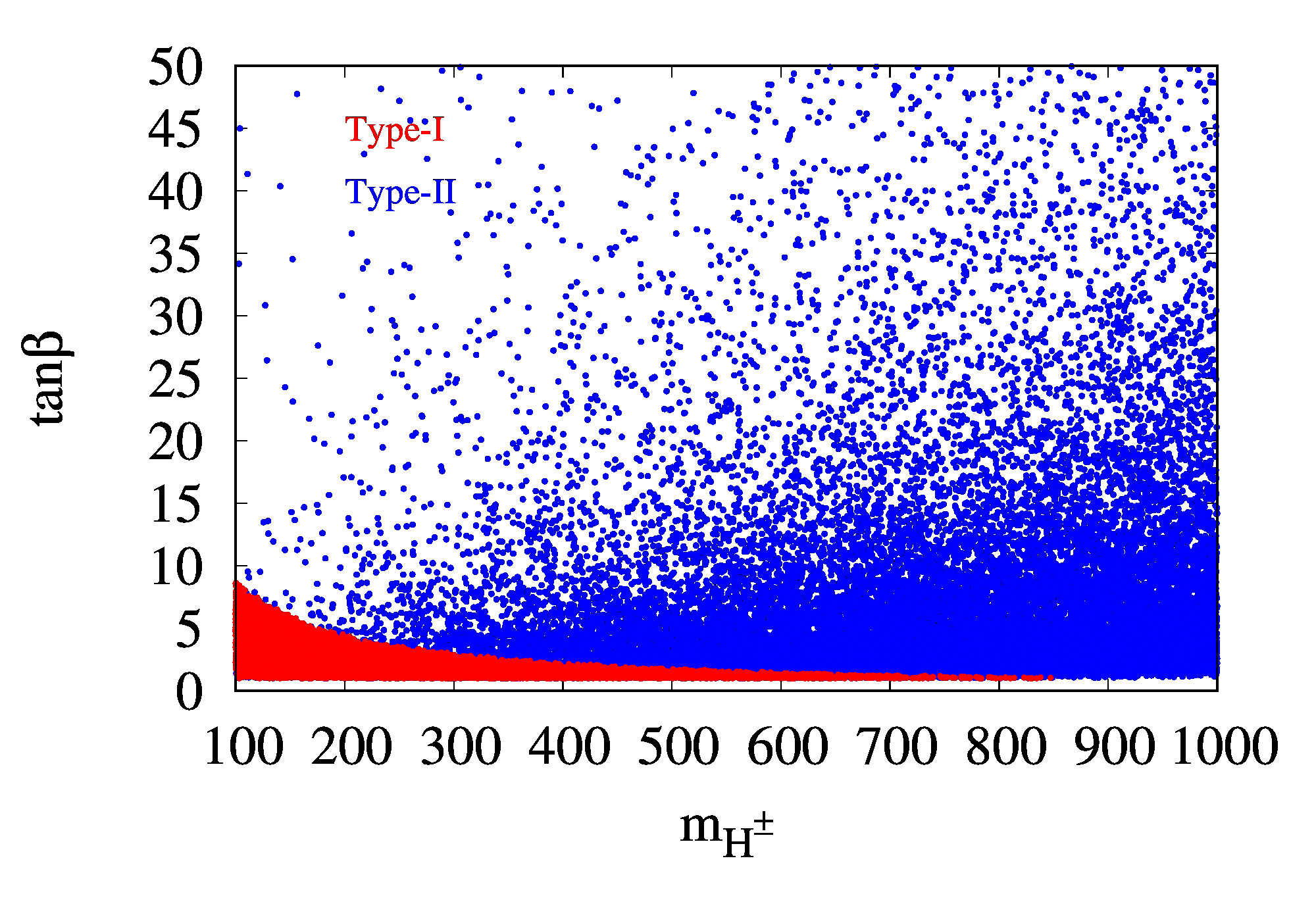}&\includegraphics[scale=0.12]{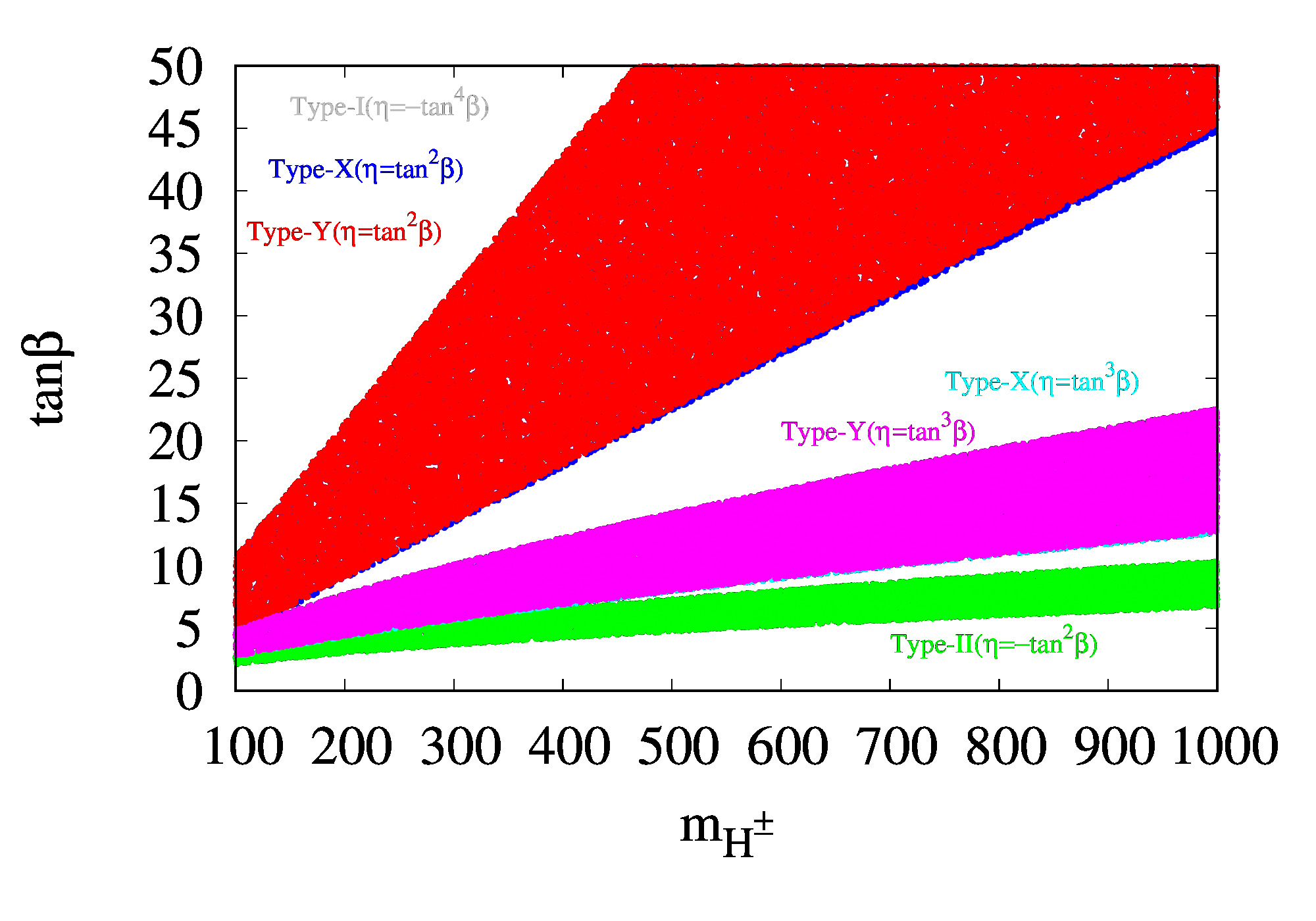}\\
(a) & (b)
\end{tabular}
\caption{\label{RtD2}
$\RtDs$-fitting results of
$\tan\beta$ vs $m_{H^\pm}$ for (a) free anomalous couplings within $-1000\le\eta\le 1000$ and 
(b) $\eta=(\pm)\tan^\alpha\beta$ for some powers of $\alpha$ for different Types of the model, at the $1\sigma$ level.
Regions for Type-I with $\eta=-\tan^4\beta$, Type-X with $\eta=\tan^2\beta$,
and Type-Y with $\eta=\tan^2\beta$ are overlapped; 
regions for Type-X with $\eta=\tan^3\beta$ and Type-Y with $\eta=\tan^3\beta$ are also overlapped.
}
\end{figure}

Figure \ref{RtD2} shows the allowed region in the $m_{H^\pm}$-$\tan\beta$ plane to fit the $\RtDs$.
%----------------- Figure 6 ------------------------------------------------
%------------------------------------------------------------------------------
In Fig.\ \ref{RtD2}(a), $\eta$ is a free parameter. 
For Types X and Y, almost the entire region is allowed.
The different behaviors of the Type I and II models are due to the factors of 
$\eta/\tan^2\beta$ (Type-I) and $\eta\tan^2\beta$ (Type-II) in Eq.\ (\ref{RDHpm2}).
\par
In Fig.\ \ref{RtD2}(b), $\eta\equiv\pm\tan^\alpha\beta$ for some $\alpha$. 
Compared to Fig.\ \ref{RD2}(b), each Type shows similar behavior, but with broader bands.
The reason is that the $\RtDs$ values are more consistent with each other than $\RDs$ ones,
and thus more points in the $m_{H^\pm}$-$\tan\beta$ plane are allowed around $\chi^2_\text{min}$.
And the bands for Types X and Y with $\eta=\tan^2\beta$ stretch to the region of $m_{H^\pm}=1000$ GeV.
%%%%%%%%%%%%%%%%%%%%%%%%%%%%%%%%%%%%%%%%%%%%%%%%%%%%%%%%%%%%%%%%%%%%%%%%%%%%%%%%
\section{Conclusions}
%%%%%%%%%%%%%%%%%%%%%%%%%%%%%%%%%%%%%%%%%%%%%%%%%%%%%%%%%%%%%%%%%%%%%%%%%%%%%%%%
In this work we tried to solve the puzzle of $\RDs$ in the 2HDM.
We introduced $\eta$ as an anomalous $\tau$ coupling to $H^+$ to fit the data through minimizing $\chi^2$.
To fit the excess of the data over the SM predictions it needs to enhance the charged Higgs contributions,
which come in the form of 
$\RDs_{H^\pm}\sim\eta m_bm_\tau/m_{H^\pm}^2+(\eta m_bm_\tau/m_{H^\pm}^2)^2$.
Thus, for small values of $\eta$, $m_{H^\pm}$ cannot be large enough to avoid detection.
For the Type-II the situation is alleviated because there is already a factor of 
$\tan^2\beta$ (but with opposite sign) in $\RDs_{H^\pm}$.
As shown in Fig.\ \ref{RD2}(b), a large $m_{H^\pm}\sim1000$ GeV is allowed if 
$\RDs_{H^\pm}\sim (m_bm_\tau/m_{H^\pm}^2)\tan^3\beta
+(m_bm_\tau/m_{H^\pm}^2)^2\tan^6\beta$ in any Type of 2HDM model.
\par
The new ratios $\RtDs$ fit much better.
Contributions of the form 
$\sim(m_bm_\tau/m_{H^\pm}^2)\tan^2\beta
+(m_bm_\tau/m_{H^\pm}^2)^2\tan^4\beta$ 
allow a large $m_{H^\pm}\sim 1000$ GeV as in Fig.\ \ref{RtD2}(b), which is not true for the $\RDs$ fitting.
In both cases of $\RDs$ and $\RtDs$ fitting, any type of 2HDM is as good as another 
with an appropriate $\eta$.
For a sufficiently large $m_{H^\pm}\gtrsim 1000$ GeV,
new contributions of the form $\sim (m_bm_\tau/m_{H^\pm}^2)\tan^k\beta$ 
with $k=2$ fit the data well for $\RtDs$, while $k\ge 3$ for $\RDs$.
It should be noted that the errors in $\RtDs$ are still large.

%%%%%%%%%%%%%%%%%%%%%%%%%%%%%%%%%%%%%%%%%%%%%%%%%%%%%%%%%%%%%%%%%%%%%%%%%%%%%%%%


\begin{thebibliography}{99}
%%%%%%%%%%%%%%%%%%%%%%%%%%%%%%%%%%%%%%%%%%%%%%%%%%%%%%%%%%%%%%%%%%%%%%%%%%%%%%%%
\bibitem{Na} 
  H.~Na {\it et al.} [HPQCD Collaboration],
  %``$B \rightarrow D l \nu$ form factors at nonzero recoil and extraction of $|V_{cb}|$,''
  Phys.\ Rev.\ D {\bf 92}, no. 5, 054510 (2015)
  Erratum: [Phys.\ Rev.\ D {\bf 93}, no. 11, 119906 (2016)]
  doi:10.1103/PhysRevD.93.119906, 10.1103/PhysRevD.92.054510
  [arXiv:1505.03925 [hep-lat]].
  %%CITATION = doi:10.1103/PhysRevD.93.119906, 10.1103/PhysRevD.92.054510;%%
\bibitem{Fajfer} 
  S.~Fajfer, J.~F.~Kamenik and I.~Nisandzic,
  %``On the $B \to D^* \tau \bar \nu_{\tau}$ Sensitivity to New Physics,''
  Phys.\ Rev.\ D {\bf 85}, 094025 (2012)
  doi:10.1103/PhysRevD.85.094025
  [arXiv:1203.2654 [hep-ph]].
  %%CITATION = doi:10.1103/PhysRevD.85.094025;%%
\bibitem{BaBar1} 
  J.~P.~Lees {\it et al.} [BaBar Collaboration],
  Phys.\ Rev.\ D {\bf 88}, no. 7, 072012 (2013)
  doi:10.1103/PhysRevD.88.072012
  [arXiv:1303.0571 [hep-ex]].
  %%CITATION = doi:10.1103/PhysRevD.88.072012;%%
  \bibitem{BaBar_PRL} 
  J.~P.~Lees {\it et al.} [BaBar Collaboration],
  %``Evidence for an excess of $\bar{B} \to D^{(*)} \tau^-\bar{\nu}_\tau$ decays,''
  Phys.\ Rev.\ Lett.\  {\bf 109}, 101802 (2012)
  doi:10.1103/PhysRevLett.109.101802
  [arXiv:1205.5442 [hep-ex]].
  %%CITATION = doi:10.1103/PhysRevLett.109.101802;%%
\bibitem{Belle1} 
  M.~Huschle {\it et al.} [Belle Collaboration],
  Phys.\ Rev.\ D {\bf 92}, no. 7, 072014 (2015)
  doi:10.1103/PhysRevD.92.072014
  [arXiv:1507.03233 [hep-ex]].
  %%CITATION = doi:10.1103/PhysRevD.92.072014;%%
%\bibitem{Belle1603}
% A.~Abdesselam {\it et al.} [Belle Collaboration],
  %``Measurement of the branching ratio of $\bar{B}^0 \rightarrow D^{*+} \tau^- \bar{\nu}_{\tau}$ relative to $\bar{B}^0 \rightarrow D^{*+} \ell^- \bar{\nu}_{\ell}$ decays with a semileptonic tagging method,''
%  arXiv:1603.06711 [hep-ex].
%%CITATION = ARXIV:1603.06711;%%
\bibitem{Belle1607} 
  Y.~Sato {\it et al.} [Belle Collaboration],
  %``Measurement of the branching ratio of $\bar{B}^0 \rightarrow D^{*+} \tau^- \bar{\nu}_{\tau}$ relative to $\bar{B}^0 \rightarrow D^{*+} \ell^- \bar{\nu}_{\ell}$ decays with a semileptonic tagging method,''
  Phys.\ Rev.\ D {\bf 94}, no. 7, 072007 (2016)
  doi:10.1103/PhysRevD.94.072007
  [arXiv:1607.07923 [hep-ex]].
  %%CITATION = doi:10.1103/PhysRevD.94.072007;%%
\bibitem{Belle1612}
S.~Hirose {\it et al.} [Belle Collaboration],
  %``Measurement of the $\tau$ lepton polarization and $R(D^*)$ in the decay $\bar{B} \to D^* \tau^- \bar{\nu}_\tau$,''
  Phys.\ Rev.\ Lett.\  {\bf 118}, no. 21, 211801 (2017)
  doi:10.1103/PhysRevLett.118.211801
  [arXiv:1612.00529 [hep-ex]].
  %%CITATION = doi:10.1103/PhysRevLett.118.211801;%%
\bibitem{LHCb1} 
  R.~Aaij {\it et al.} [LHCb Collaboration],
  Phys.\ Rev.\ Lett.\  {\bf 115}, no. 11, 111803 (2015)
  Addendum: [Phys.\ Rev.\ Lett.\  {\bf 115}, no. 15, 159901 (2015)]
  doi:10.1103/PhysRevLett.115.159901, 10.1103/PhysRevLett.115.111803
  [arXiv:1506.08614 [hep-ex]].
%%CITATION = doi:10.1103/PhysRevLett.115.159901, 10.1103/PhysRevLett.115.111803;%%  
%----------------------------------------------------------------------------------------------
\bibitem{Andreas}
A.~Crivellin, C.~Greub and A.~Kokulu,
  %``Explaining $B\to D\tau\nu$, $B\to D^*\tau\nu$ and $B\to \tau\nu$ in a 2HDM of type III,''
  Phys.\ Rev.\ D {\bf 86}, 054014 (2012)
  doi:10.1103/PhysRevD.86.054014
  [arXiv:1206.2634 [hep-ph]];
  %%CITATION = doi:10.1103/PhysRevD.86.054014;%%
  %``Flavor-phenomenology of two-Higgs-doublet models with generic Yukawa structure,''
  Phys.\ Rev.\ D {\bf 87}, no. 9, 094031 (2013)
  doi:10.1103/PhysRevD.87.094031
  [arXiv:1303.5877 [hep-ph]];
  %%CITATION = doi:10.1103/PhysRevD.87.094031;%%
A.~Crivellin, J.~Heeck and P.~Stoffer,
  %``A perturbed lepton-specific two-Higgs-doublet model facing experimental hints for physics beyond the Standard Model,''
  Phys.\ Rev.\ Lett.\  {\bf 116}, no. 8, 081801 (2016)
  doi:10.1103/PhysRevLett.116.081801
  [arXiv:1507.07567 [hep-ph]].
  %%CITATION = doi:10.1103/PhysRevLett.116.081801;%%  
\bibitem{Fazio}
P.~Biancofiore, P.~Colangelo and F.~De Fazio,
  %``On the anomalous enhancement observed in $B \to D^{(*)}\tau{\bar \nu}_\tau$ decays,''
  Phys.\ Rev.\ D {\bf 87}, no. 7, 074010 (2013)
  doi:10.1103/PhysRevD.87.074010
  [arXiv:1302.1042 [hep-ph]].
  %%CITATION = doi:10.1103/PhysRevD.87.074010;%%
\bibitem{Cline}
J.~M.~Cline,
  %``Scalar doublet models confront ? and b anomalies,''
  Phys.\ Rev.\ D {\bf 93}, no. 7, 075017 (2016)
  doi:10.1103/PhysRevD.93.075017
  [arXiv:1512.02210 [hep-ph]].
  %%CITATION = doi:10.1103/PhysRevD.93.075017;%%
\bibitem{Koerner}
M.~A.~Ivanov, J.~G.~K\"orner and C.~T.~Tran,
 %``Analyzing new physics in the decays $\bar{B}^0 \to D^{(\ast)}\tau^-\bar\nu_{\tau}$ with form factors obtained from the covariant quark model,''
  Phys.\ Rev.\ D {\bf 94}, no. 9, 094028 (2016)
  doi:10.1103/PhysRevD.94.094028
  [arXiv:1607.02932 [hep-ph]];
  %%CITATION = doi:10.1103/PhysRevD.94.094028;%%
 %``Probing new physics in $\bar{B}^0 \to D^{(\ast)} \tau^- \bar\nu_{\tau}$ using the longitudinal, transverse, and normal polarization components of the tau lepton,''
  Phys.\ Rev.\ D {\bf 95}, no. 3, 036021 (2017)
  doi:10.1103/PhysRevD.95.036021
  [arXiv:1701.02937 [hep-ph]].
  %%CITATION = doi:10.1103/PhysRevD.95.036021;%%
%----------------------------------------------------------------------------------------------
\bibitem{Dhargyal} 
  L.~Dhargyal,
  %``$R(D^{(*)})$ and $\mathcal{B}r(B \rightarrow \tau\nu_{\tau})$ in a Flipped/Lepton-Specific 2HDM with anomalously enhanced charged Higgs coupling to $\tau$/b,''
  Phys.\ Rev.\ D {\bf 93}, no. 11, 115009 (2016)
  doi:10.1103/PhysRevD.93.115009
  [arXiv:1605.02794 [hep-ph]].
  %%CITATION = doi:10.1103/PhysRevD.93.115009;%%  
\bibitem{Nandi} 
S.~Nandi, S.~K.~Patra and A.~Soni,
 %``Correlating new physics signals in $B \to D^{(*)} \tau \nu_{\tau}$ with $B \to \tau \nu_{\tau}$,''
arXiv:1605.07191 [hep-ph].
%%CITATION = ARXIV:1605.07191;%%
\bibitem{Aoki}
M.~Aoki, S.~Kanemura, K.~Tsumura and K.~Yagyu,
  %``Models of Yukawa interaction in the two Higgs doublet model, and their collider phenomenology,''
  Phys.\ Rev.\ D {\bf 80}, 015017 (2009)
  doi:10.1103/PhysRevD.80.015017
  [arXiv:0902.4665 [hep-ph]].
%%CITATION = doi:10.1103/PhysRevD.80.015017;%%
\bibitem{Archer}
P.~R.~Archer,
  %``The Fermion Mass Hierarchy in Models with Warped Extra Dimensions and a Bulk Higgs,''
  JHEP {\bf 1209}, 095 (2012)
  doi:10.1007/JHEP09(2012)095
  [arXiv:1204.4730 [hep-ph]].
  %%CITATION = doi:10.1007/JHEP09(2012)095;%%
\bibitem{Agashe}
K.~Agashe, T.~Okui and R.~Sundrum,
  %``A Common Origin for Neutrino Anarchy and Charged Hierarchies,''
  Phys.\ Rev.\ Lett.\  {\bf 102}, 101801 (2009)
  doi:10.1103/PhysRevLett.102.101801
  [arXiv:0810.1277 [hep-ph]].
  %%CITATION = doi:10.1103/PhysRevLett.102.101801;%%  
\bibitem{BaBar2}
 J.~P.~Lees {\it et al.} [BaBar Collaboration],
  %``Evidence of $B^+ \to \tau^+\nu$ decays with hadronic B tags,''
  Phys.\ Rev.\ D {\bf 88}, no. 3, 031102 (2013)
  doi:10.1103/PhysRevD.88.031102
  [arXiv:1207.0698 [hep-ex]].
  %%CITATION = doi:10.1103/PhysRevD.88.031102;%%
\bibitem{Belle3}
 B.~Kronenbitter {\it et al.} [Belle Collaboration],
  %``Measurement of the branching fraction of B^+ -> tau^+ nu_tau decays with the semileptonic tagging method,''
  Phys.\ Rev.\ D {\bf 92}, no. 5, 051102 (2015)
  doi:10.1103/PhysRevD.92.051102
  [arXiv:1503.05613 [hep-ex]].
  %%CITATION = doi:10.1103/PhysRevD.92.051102;%%
%
\end{thebibliography}
\end{document}